# Highly Accelerated EPI with Wave Encoding and Multi-shot Simultaneous Multi-Slice Imaging

Jaejin Cho[1,2*], Congyu Liao[3], Qiyuan Tian[1,2], Zijing Zhang[4], Jinmin Xu[4], Wei-Ching Lo[5], Benedikt A. Poser[6], V. Andrew Stenger[7], Jason Stockmann[1,2,8], Kawin Setsompop[3], Berkin Bilgic[1,2,8]

1. Athinoula A. Martinos Center for Biomedical Imaging, Massachusetts General Hospital, Boston, MA, United States
2. Harvard Medical School, Boston, MA, United States
3. Radiological Sciences Laboratory, Stanford University, Palo Alto, CA, United States
4. State Key Laboratory of Modern Optical Instrumentation, College of Optical Science and Engineering, Zhejiang University, Hangzhou, China
5. Siemens Medical Solutions, Boston, MA, United States
6. Maastricht Brain Imaging Centre, Maastricht University, Maastricht, the Netherlands
7. MR Research Program, Department of Medicine, John A. Burns School of Medicine, University of Hawai'i, HI, USA
8. Harvard-MIT Health Sciences and Technology, Massachusetts Institute of Technology, Cambridge, MA, United States

**Correspondence to**: Jaejin Cho, Ph.D.

**Running head:** Wave-EPI for Multi-shot SMS Imaging

## *Abstract*

*Purpose*: To introduce wave encoded acquisition and reconstruction techniques for highly accelerated echo planar imaging (EPI) with reduced g-factor penalty and image artifacts.

*Method*: Wave-EPI involves playing sinusoidal gradients during the EPI readout while employing interslice shifts as in blipped-CAIPI acquisitions. This spreads the aliasing in all spatial directions, thereby taking better advantage of 3D coil sensitivity profiles. The amount of voxel spreading that can be achieved by the wave gradients during the short EPI readout period is constrained by the slew rate of the gradient coils and peripheral nerve stimulation (PNS) monitor. We propose to use a "half-cycle" sinusoidal gradient to increase the amount of voxel spreading that can be achieved while respecting the slew and stimulation constraints. Extending wave-EPI to multi-shot acquisition minimizes geometric distortion and voxel blurring at high in-plane resolution, while structured low-rank regularization mitigates shot-to-shot phase variations without additional navigators. We propose to use different point spread functions (PSFs) for the k-space lines with positive and negative polarities, which are calibrated with a FLEET-based reference scan and allow for addressing gradient imperfections.

*Results*: Wave-EPI provided whole-brain single-shot gradient echo (GE) and multi-shot spin echo (SE) EPI acquisitions at high acceleration factors and was combined with g-Slider slab encoding to boost the SNR level in 1mm isotropic diffusion imaging. Relative to blipped-CAIPI, wave-EPI reduced average and maximum g-factors by up to 1.21- and 1.37-fold, respectively.

*Conclusion*: Wave-EPI allows highly accelerated single- and multi-shot EPI with reduced g-factor and artifacts and may facilitate clinical and neuroscientific applications of EPI by improving the spatial and temporal resolution in functional and diffusion imaging.

*Keywords:* Diffusion imaging, functional imaging, g-Slider, Low-rank reconstruction, Multi-shot EPI, Parallel imaging, SMS imaging, Wave-CAIPI, Wave-EPI

# *Introduction*

Echo planar imaging (EPI) provides fast encoding per imaging slice and has found wide application in clinical and neuroscientific acquisitions (1–3), especially in functional (fMRI) and diffusion (dMRI) imaging (4,5). However, single-shot (SS-) EPI suffers from severe susceptibility-induced geometric distortion and $T_2$- and $T_2^*$-induced voxel blurring due to low bandwidth along the phase-encoding direction (6,7). These artifacts only get worse at higher in-plane resolution as the time it takes to acquire each line of k-space increases approximately linearly.

In-plane parallel imaging (PI) acceleration is an effective approach to mitigate EPI-related artifacts by reducing the effective echo-spacing (8–10). Existing PI approaches aim to achieve high accelerations by exploiting linear dependencies across the multi-coil data, either through interpolation across k-space (e.g. GRAPPA, 11), or image-space reconstruction using a pre-calculated coil sensitivity map (e.g. SENSE, 12) More recent work, ESPIRiT, has bridged the gap between GRAPPA and SENSE (13).

Another approach for mitigating EPI-related artifacts is multi-shot (MS-) acquisition. This also provides a reduction in effective echo spacing by segmenting the k-space into multiple portions which are covered across multiple repetition times (TRs). However, potential shot-to-shot phase variations across multiple EPI shots may lead to additional artifacts in shot-combined images (14–16). Recent algorithms including multiplexed sensitivity encoding (MUSE, 17) and multi-shot sensitivity encoded diffusion data recovery using structured low-rank matrix completion (MUSSELS, 18) allow for successful and navigator-free combination of these shots. MUSE achieves this by estimating phase variations from interim shot images and incorporating these into a final joint SENSE reconstruction where all the shot data are combined. MUSSELS jointly reconstructs MS-EPI images using Hankel structured low-rank constraint across the EPI shots. This structured low-rank prior has also been explored in SAKE (19), ALOHA (20) and LORAKS (21,22) for image reconstruction in spin-warp acquisitions as well as Nyquist ghost correction (23,24) in EPI.

Simultaneous multislice (SMS) acquisition excites and encodes multiple imaging slices simultaneously to offer a significant reduction in TR. Controlled aliasing in parallel imaging (CAIPI) modulates the phase of excited slices in k-space to improve the PI conditioning for resolving collapsed slice data (25). This phase modulation induces inter-slice shifts in the phase-encoding (PE) direction so that image aliasing is distributed in both PE and slice dimensions. Blipped-CAIPI extends the CAIPI strategy to echo-planar trajectories (26). This modulates the phase of the excited slices by using small gradient-blips in the slice-selection direction during ramp-time between the adjacent readouts to create inter-slice shifts in the PE direction.

Wave-CAIPI is a more recent controlled aliasing method that can further reduce noise amplification and aliasing artifacts (27,28). It employs extra sinusoidal gradient modulations in the phase- and the partition-encoding directions during the readout to better harness coil sensitivity variations in all three-dimensions. Wave encoding also incorporates 2D-CAIPI inter-slice shifts to improve the PI conditioning (29), and has found applications in highly accelerated gradient echo (GE), MPRAGE, and fast spin echo acquisitions (28,30–33).

We had previously combined wave encoding with EPI readout for GE contrast and demonstrated high in-plane acceleration to reduce geometric distortion and $T_2^*$ blurring in earlier ISMRM proceedings (34–37). In this work, we report improvements in calibration, acquisition and reconstruction strategies for wave-EPI. Specifically,

1. We use different point spread functions (PSFs) to represent the voxel spreading effect of sinusoidal gradients for the k-space lines with positive and negative polarities. This "dual PSF" approach allows us to address potential imperfections in all three gradient axes. We estimate these PSFs from rapid calibration scans with FLEET acquisition ordering (38).
2. The amount of voxel spreading that can be achieved by the wave gradients during the short EPI readout is constrained by the slew rate and the peripheral nerve stimulation limit. We propose to increase the amount of voxel spreading using a "half-cycle" sinusoidal.
3. We incorporate SMS encoding to improve efficiency, and MS capability to further reduce distortion and $T_2/T_2^*$-induced voxel blurring. We extend these capabilities to spin-echo (SE) and diffusion contrasts. Incorporating structured low-rank regularization allows for leveraging the similarities across the shots for improved reconstruction and accounts for potential shot-to-shot phase differences.
4. We take advantage of the SNR benefit provided by the volumetric noise averaging in the RF-encoded g-Slider approach (39), and push the spatial resolution in wave-EPI acquisition. We propose a joint image reconstruction method that incorporates g-Slider slab encoding operator into the wave-EPI forward model to minimize potential resolution loss due to an intravoxel blurring effect from wave gradients.

These improvements allowed us to demonstrate up to 1.21- and 1.37-fold gain in average and maximum g-factor noise amplification over blipped-CAIPI for single-shot GE-EPI. Wave-EPI provided marked improvement in image quality at the high acceleration rate of $R_{in}xR_{sms}$=5x2-fold *per shot* in diffusion acquisition using two shots of EPI. Further, we demonstrate high-resolution diffusion data at 1 mm isotropic voxel size by combining g-Slider with wave-EPI at $R_{in}xR_{sms}$=6x2-fold acceleration using two EPI shots with high SNR and geometric fidelity and improved g-factor mitigation.

## *Theory*

### *Voxel Spreading by Wave-Encoding*

Wave-encoding employs additional sinusoidal gradients during the readout to take better advantage of 3D coil sensitivity profiles (27). Wave-encoding modulates the phase during the readout and incurs a corkscrew trajectory in k-space. The wave-encoded signal, $s$, can be explained using the following equation.

$$s(t) \;=\; \int_{x,y,z} m(x,y,z) e^{-i2\pi(k_x(t)x + k_y y + k_z z)} e^{-i\gamma \int_0^t (g_y(\tau)y + g_z(\tau)z)d\tau} dxdydz \qquad [1]$$

where $m$ is the magnetization, $\gamma$ is the gyromagnetic ratio, and $g$ is the time-varying sinusoidal wave gradient. Voxel spreading by wave-encoding, $d$, can be obtained from the equation of wave-encoded signal as follows.

$$s(t) \;=\; \int_{x,y,z} m(x,y,z) e^{-i2\pi(k_x(t)(x + d(t,y,z)) + k_y y + k_z z)} dxdydz \qquad [2]$$

$$d(t,y,z) \;=\; \frac{\gamma \int_0^t (g_y(\tau)y + g_z(\tau)z)d\tau}{2\pi k_x(t)} \qquad [3]$$

Voxel spreading by wave-encoding is a function of readout time and the position in the y- and z-dimensions. Because this is continually changing with the readout time, voxel signal is widely spread in the readout direction, which spreads aliasing in three-dimensions in accelerated acquisitions.

### *Limited Voxel Spreading in EPI Acquisition*

In EPI acquisition, the maximum amount of signal displacement is limited by the short readout window. To explain the relation between signal displacement and the readout window, we first describe the signal displacement by cosine wave-encoding in the y-direction as follows.

$$
\begin{aligned}
d(t,y) &= y \cdot \frac{\gamma \int_0^t g_y(\tau)d\tau}{2\pi k_x(t)} \\
&= y \cdot \frac{1}{2\pi} \cdot \frac{\int_0^t G_w \cos\left(2\pi \frac{n_c}{T_r}\tau\right) d\tau}{\int_0^t G_x \tau d\tau} \\
&= y \cdot \frac{1}{2\pi} \cdot \frac{G_w}{G_x} \cdot \frac{\sin\left(2\pi \frac{n_c}{T_r} t\right)}{2\pi \frac{n_c}{T_r} t} \\
&= y \cdot \frac{1}{2\pi} \cdot \frac{G_w}{G_x} \cdot \text{sinc}\left(2\pi \frac{n_c}{T_r} t\right) \quad [4]
\end{aligned}
$$

where $T_r$ is the duration of the readout window, $n_c$ is the number of wave cycles, $G_w$ is the amplitude of the cosine wave-encoding gradient, and $G_x$ is the amplitude of the readout gradient. The maximum of $G_w$ is constrained by the slew rate of the gradient coil, $R_{max}$, due to:

$$
G_w \leq \frac{R_{max} T_r}{2\pi n_c} \quad [5]
$$

The voxel spreading can be described using the $R_{max}$, $T_r$, and $n_c$ as follows.

$$
d(t,y) \leq y \cdot \frac{1}{2\pi} \cdot \frac{1}{G_x} \cdot \frac{R_{max} T_r}{2\pi n_c} \cdot \text{sinc}\left(2\pi \frac{n_c}{T_r} t\right) \quad [6]
$$

Increasing $T_r$ can maximize voxel spreading and improve conditioning of parallel imaging reconstruction. However, in EPI acquisition, high receiver bandwidth is required to mitigate EPI-related artifacts by reducing echo-spacing time, which limits the achievable voxel spreading. The spreading incurred by the sine wave-encoding in the z-direction is similarly limited. In equation [6], most of the variables are constrained by FOV, hardware specifications and receiver bandwidth; only $n_c$ can be adjusted. In this paper, we increase the maximum voxel spreading by minimizing $n_c$ while maintaining a high receiver bandwidth in EPI acquisition. We chose a "half-cycle" cosine and one cycle sine wave-encoding in the y- and z- directions, respectively.

## Methods

### Wave-EPI

Wave-CAIPI strategy was applied to EPI readout to mitigate geometric distortion and $T_2$ or $T_2^*$ blurring by minimizing the effective echo-spacing with high in-plane undersampling and reduce the TR using multiband acceleration. Figure 1a shows the proposed wave-SE-EPI sequence diagram, where "half-cycle" of cosine wave gradients and one-cycle of sine wave gradients were applied in the phase-encoding and the slice-selection directions, respectively. This creates the usual corkscrew wave trajectory as shown in Figure 1b, but doubles the amplitude in the phase-encoding sinusoidal in the presence of stringent slew-rate limitations, provided that the PNS limit is not exceeded. Because the polarity of the readout gradient is changing across even and odd EPI readouts, the polarity of the sinusoidal wave-gradients is also flipped to maintain the consistency of the k-space trajectory of the wave-EPI sequence.

### Dual PSF and Calibration scan

Because eddy currents and system imperfections can create differences between the actual and theoretical wave-PSFs, accurate characterization of the actual k-space trajectory is needed. In addition, wave-encoding gradients are continuously flipping their polarity to maintain the trajectory consistency in the k-space, as a result, the positive and negative readouts may have different PSFs due to such imperfections. We propose to calibrate and use "dual" wave-PSFs to de-convolve the positive and negative readouts as shown in Figure 1c.

To estimate the experimental PSF, we developed a reference scan based on the fast low-angle excitation echo-planar technique (FLEET) with and without wave-encoding (38). Reference scans were acquired twice with opposite readout gradient-polarities (40) to estimate PSFs in both positive and negative polarities. PSFs can be directly calculated from the phase difference between calibration data with and without wave-encoding in the $k_x$-y or $k_x$-z hybrid domain. However, due to low-flip angle excitation, this direct approach may suffer from low SNR, especially in the high-frequency region. Therefore, we use a sparse frequency approach called "auto-PSF" to estimate PSFs in the high $k_x$ regions (41). The auto-PSF technique assumes that the wave-encoding gradient has sparse frequency content in the Fourier-transformed domain due to its sinusoidal shape, therefore, only a small number of frequency variables are needed to represent wave-encoding. Instead of fitting the entire wave-encoding trajectory, we looked for the sparse frequency variables to estimate the wave-PSF as follows.

$$\varrho = \underset{\varrho}{\operatorname{argmin}} \left( \left\| \mathcal{F}_y P(\varrho) \mathcal{F}_y^{-1} S_r - S_w \right\|_2^2 \right) \qquad [7]$$

where $\varrho$ is the sparse frequency variables representing wave-encoding in its Fourier-transformed domain, $P$ is the wave-PSF generator using the sparse frequency variables, $S_r$ is the k-space reference signal without wave-encoding, and $S_w$ is the k-space reference signal with wave-encoding, respectively. In this paper, $\varrho$ only has the four complex numbers to present the wave-encoding. We separately estimated $\varrho$ for each positive and negative EPI readout as well as each gradient in the y- and z-direction. As a result, four different PSFs are calibrated as shown in Figure 1c.

### *Wave-EPI Image Reconstruction*

We reconstructed single-shot wave-EPI data using dual wave-PSFs and coil sensitivity maps calibrated from additional reference scans as follows.

$$I = \underset{I}{\operatorname{argmin}} \left\| \begin{bmatrix} W_p \mathcal{F}_y P(\varrho_p) \\ W_n \mathcal{F}_y P(\varrho_n) \end{bmatrix} \mathcal{F}_x CI - \begin{bmatrix} S_p \\ S_n \end{bmatrix} \right\|_2^2 \qquad [8]$$

where $I$ is the image, $p$ denotes the positive lines, $n$ denotes the negative lines, $C$ is the coil-sensitivity map, and $S_p$ and $S_n$ are the acquired wave-EPI k-space signal from the positive and negative readouts, respectively. We extend the single-shot wave-EPI reconstruction to multi-shot wave-EPI by exploiting low-rankness across multiple EPI shots using the MUSSELS approach. The reconstruction of the wave-epi with Hankel low-rank constraint between multiple shots is described as follows.

$$I = \underset{I}{\operatorname{argmin}} \sum_s \left\| \begin{bmatrix} W_p \mathcal{F}_y P(\varrho_p) \\ W_n \mathcal{F}_y P(\varrho_n) \end{bmatrix} \mathcal{F}_x CI_s - \begin{bmatrix} S_{s,p} \\ S_{s,n} \end{bmatrix} \right\|_2^2 + \|\mathcal{H}(I)\|_* \qquad [9]$$

where $I_s$ is the reconstructed image of the $s$-th shot, $\mathcal{H}$ is the Hankel matrix, and $\|\cdot\|_*$ represents nuclear norm constraints, respectively. We exploit fast iterative shrinkage-thresholding algorithm (FISTA, 40) to accelerate the reconstruction speed. To minimize the nuclear norm constraints, we employed the SVD-based thresholding approach (19,21).

Exemplar code and data can be found at "*https://github.com/jaejin-cho/wave-EPI*"

*Joint Image Reconstruction for Wave-EPI with g-Slider RF-encoding*

Wave-EPI is also able to take advantage of the g-Slider approach which provides SNR gain by performing multiple RF encoded acquisitions over a thick slab to resolve thin slice information (39). However, equation [9] may induce a nonnegligible voxel blurring due to more intra-voxel phase variation within an e.g. 5 mm thick slab. We used wave-PSFs at the thin-slice resolution to combat this, and directly estimate the thin-slice resolution images. We propose a joint image reconstruction method to take advantage of g-Slider's SNR gain while minimizing voxel blurring caused by wave-encoding. We estimate shot-to-shot phase variations from the reconstructed interim RF-encoded and multi-shot images using the equation [9] in advance, and use this information during the joint reconstruction. This joint reconstruction using the pre-estimated shot-to-shot phase information for wave-EPI with g-Slider RF-encoding to reconstruct the thin-slice resolution image $I_h$ can be described as follows.

$$I_h = \underset{I_h}{\mathrm{argmin}} \sum_s \sum_g \left\| \begin{bmatrix} W_p \mathcal{F}_y G_r P(\varrho_p) \\ W_n \mathcal{F}_y G_r P(\varrho_n) \end{bmatrix} \mathcal{F}_x C \Phi_{s,r} I_h - \begin{bmatrix} S_{s,r,p} \\ S_{s,r,n} \end{bmatrix} \right\|_2^2 \quad [10]$$

where $r$ denotes g-Slider RF encoding index, $G_r$ is the RF encoding matrix, $\Phi_{s,r}$ is the estimated shot-to-shot phase variation in shot $s$, RF index $r$, and $S_{s,r,p}$ and $S_{s,r,n}$ are the acquired $s$-th shot, $r$-th RF encoding signal from positive and negative readouts, respectively. To provide an initial guess to the optimizer, we directly applied $G_r^{-1}$ to the phase-removed interim shot images.

*Data Acquisition*

Three different in-vivo experiments highlighting the versatility of wave-EPI were conducted on a 3T Siemens Prisma system with a 32-channel head coil. Table 1 shows the imaging parameters used for the in-vivo experiments; single-shot GE-EPI at two different reduction factors, two-shot SE-EPI for dMRI, and two-shot SE-EPI with g-Slider for dMRI (39). For MS-EPI, we present the reduction factor per single EPI shot ($R_{shot}=R_{in}$x$R_{sms}$). Blipped-CAIPI was used for comparison in all experiments. Coil sensitivity maps were calculated from pre-acquired low-resolution FLASH scans using ESPIRiT (13,43). For dMRI, we used 1000s/mm$^2$ of b-value and employed Hankel structured low-rank regularization with 7x7 kernels to mitigate the shot-to-shot phase variations. We kept 37.5% of singular values during the SVD-truncation for the low-rank regularization, where we estimated the truncation level by maximizing the 2nd fiber population count as shown in Supporting Figure S1 (44,45). We used the DWI-denoising algorithm as implemented in the MRTRIX toolbox to further denoise the reconstructed images (46,47) and FSL diffusion toolbox using tract-based spatial statistics for diffusion processing (48). Five different RF pulses were used for g-Slider acquisition to encode thin-slice information, which resolve a 5-mm slice into five 1-mm slices (39). For GE-EPI and SE-EPI experiments at 1.25 mm isotropic voxel resolution, we could almost achieve the maximum wave amount constrained by hardware slew-rate limitation. However, for dMRI with g-Slider experiments, we could use only 22mT/m x 19mT/m wave gradient limited by the PNS safety limitation, where the maximum wave gradient constrained by the slew-rate limitation is 40mT/m x 20mT/m.

## *Results*

### *Single-shot GE-EPI*

Figure 2 shows 1.25mm isotropic voxel size whole-brain single-shot GE-EPI images at $R_{in}xR_{sms}$=3x3. As pointed by the red arrows, blipped-CAIPI suffers from reconstruction artifacts due to high acceleration, which were mitigated by wave-encoding. Based on the g-factor analysis, wave-EPI reduced average and maximum g-factors by 1.21- and 1.37-fold compared to standard blipped-CAIPI in the center slice, respectively. tSNR maps were also calculated from a time-series comprising 50 TRs to quantify the tSNR gain of wave-EPI over blipped-CAIPI.

We further pushed the reduction factor to $R_{in}xR_{sms}$=4x3 as shown in Supporting Figure S2. Wave-EPI significantly mitigated the aliasing artifact and noise amplification for this highly accelerated scan. In the g-factor analysis, wave-EPI reduced average and maximum g-factors by 1.41- and 1.77-fold compared to the standard blipped-CAIPI and tSNR maps also show the tSNR gain of wave-EPI.

### *Two-shot SE-EPI for dMRI*

Figures 3 and 4 show average DWI images and fractional anisotropy (FA) map color-encoded by the primary eigenvector at b=1000s/mm$^2$ with 32 directions at 1.25mm isotropic voxel size using two-shot SE-EPI with $R_{in}xR_{sms}$=5x2 per shot. Wave-CAIPI was able to substantially reduce artifacts and noise amplification over blipped-CAIPI despite the high acceleration level. Supporting Figure S3 shows the DWI of single diffusion direction, which demonstrates the substantial noise and aliasing reductions in wave-EPI.

### *Two-shot SE-EPI with g-Slider for dMRI*

Figures 5 and 6 show the average DWI images and colored FA map at b=1000s/mm$^2$ for 63-direction data at 1mm isotropic voxel size using two-shot SE-EPI with $R_{in} \times R_{sms}$=6x2 per shot. Blipped-CAIPI with g-Slider suffers from reduced SNR and stripe artifact in both DWI and FA images. In Figure 5, the red arrows point to the remaining aliasing artifacts in blipped-CAIPI. Standard wave-EPI with g-Slider was able to largely mitigate both aliasing and striping artifacts, and joint reconstruction of wave-EPI and g-Slider further removed the striping artifact. In Figure 6, wave-EPI significantly improves the FA map and reduces noise. In the ROI in the solid-line box in the axial view, wave-EPI clearly visualizes sub-cortical white matter fascicles coherently fanning into the cortex and the orthogonality between the primary fiber orientations in the cerebral cortex and the cortical surface while blipped-CAIPI has difficulty representing the orthogonality. In the ROI in the dashed-line box in the axial view, wave-EPI displays in exquisite detail the gray matter bridges that span the internal capsule, giving rise to the characteristic stripes seen in the striatum. We observed the significant improvement of wave-EPI in the coronal and sagittal views as well.

## *Discussion*

We introduced wave-EPI, which extended wave encoding to EPI acquisition and incorporated inter-slice shifts, and for multi-shot acquisitions, structured low-rank regularization for navigator-free reconstruction. This extreme controlled aliasing approach was able to substantially reduce g-factor noise amplification and aliasing artifacts over the standard blipped-CAIPI technique for a range of image contrasts at high acceleration rates, with either single- or multi-shot acquisition.

Using estimated dual PSFs instead of the theoretical waveforms provided robustness against gradient imperfections in the wave-EPI reconstruction. Our FLEET-based calibration acquisitions for PSF estimation require ~1 minute of additional scan time for a 30-slab g-Slider acquisition. Diffusion and functional imaging experiments often entail the acquisition of a large number of TRs for dense sampling of q-space in dMRI, and for characterization of the hemodynamic response in fMRI. Since these acquisitions usually last several minutes, we anticipate that the additional ~1 minute of calibration will not impact their feasibility significantly. Standard blipped-CAIPI acquisitions lend themselves to GRAPPA / slice-GRAPPA based reconstructions, which also entail additional calibration scans with FLEET ordering and require similar amounts of extra acquisition time. Through our generalized-SENSE reconstruction in wave-EPI, we obviated the need for these kernel calibration acquisitions and estimated coil sensitivities from a quick, ~11 second FLASH scan for 30 slices. To calibrate the y and z wave-PSFs, the projected signal along to the z and y axes ($k_x$-z image and $k_x$-y image) can be used, respectively (49). By using the projected signal, the additional scan time can be theoretically reduced to about 7 sec.

Relatively low readout bandwidths in spin-warp applications of wave encoding allow for near-perfect g-factor performance at e.g. R=3x3 acceleration (27,28). The much shorter readout window in EPI acquisition precluded our ability to achieve similar g-factor performance. As detailed in the Theory section, a shorter readout window decreases the voxel spreading. In addition, maximum wave gradient is limited by hardware slew-rate limitation and peripheral nerve stimulation. To combat this issue, we proposed to use a "half-cycle" cosine waveform to achieve a larger gradient amplitude and voxel spreading effect while satisfying the slew-rate and stimulation constraints. We simulated the voxel spreading for half- and one-cycle waveforms with the same maximum slew-rate, at 1.25mm isotropic resolution with $R_{in}xR_{sms}$=4x3 as shown in Figure 7. One- and half-cycle cosine wave-gradients had 15mT/m and 30mT/m amplitudes in the phase-encoding direction, respectively. 15mT/m of one-cycle sine wave-gradient was applied in the slice-selection direction in both cases. Relative to one-cycle cosine wave, half-cycle cosine yielded more spreading in the image domain, which led to reduced g-factor penalty. Compared to blipped-CAIPI, wave encoding with one-cycle cosine reduced the average g-factor by 1.10-fold. With the half-cycle approach, the average g-factor gain was 1.23-fold. The g-factor gain with the half-cycle approach implies that wave-EPI provides a mean SNR benefit of ~1.5-averages of blipped-CAIPI acquisition for free. In addition, the relative stimulation levels for one TR were computed by SIEMENS MR IDEA software for pulse sequence programming as shown in Figure 7. The half-cycle cosine wave remained within PNS safety limits and provided more g-factor gain compared to the one-cycle cosine wave, which could not be played on the scanner as it exceeded stimulation limits. One-cycle of cosine wave starts and ends its gradient at the highest points, which might increase the ramp-up and -down time due to the slew-rate and stimulation limitations. On the other hand, half-cycle cosine encoding starts and ends its gradient at zero amplitude and saves its slew-rate and stimulation during the ramp-up and –down time.

In our current acquisitions, we have only sampled data during the "flat-top" portion of the EPI readout. Employing ramp sampling would further increase the acquisition efficiency and decrease the echo spacing. Our simulations indicate that the echo-spacing can be reduced by ~10% at 1mm resolution. We think that the proposed PSF calibration scan is flexible enough to allow for incorporating ramp sampling for this additional efficiency gain. Another minor drawback of wave encoding is the fact that the pre-phasing lobe of sinusoidal gradient waveform is increasing the TE by about 1ms. Assuming a typical $T_2$ value of 60 ms, this would reduce the SNR level in dMRI by ~2%. We think that wave-EPI's ability to provide higher in-plane acceleration should more than account for this difference. We anticipate that advanced hardware systems with higher slew and relaxed nerve stimulation limits (e.g. with head-insert or head-only gradients) will allow for yet higher acceleration by enabling the use of larger wave gradient amplitudes.

The voxel spreading effect of wave gradients is continuous across space, and thus has the potential to cause within-voxel blurring since the top and bottom edges of each voxel will experience slightly different PSFs. We simulated the amount of this within-voxel blurring at 1x1x1 mm$^3$ resolution in Figure 8. As shown in Figure 8c, standard wave-EPI extended full-width half-maximum (FWHM) by 0.01 mm and generated 8% of sidelobes due to the 5mm thick slab before g-Slider reconstruction. The remaining intravoxel phase variation also leads to the stripe artifacts which was observed in Figure 5. Meanwhile, joint reconstruction of wave-EPI and g-Slider only extended FWHM by 0.01 mm, which generates negligible intra-voxel blurring.

Wave-EPI reconstruction for single-shot acquisition requires 29 seconds per slice-group, whereas this is 28 seconds/slice-group in blipped-CAIPI. The additional low-rank thresholding step in multi-shot acquisition further increases the reconstruction time to 3.8 minutes/slice-group for wave- and 3.2 minutes/slice-group in blipped-CAIPI. Our current Matlab implementation makes use of parallel processing across the slice groups, but further speed-up is possible through additional parallelization across volumes, code optimization, and alternate reconstruction algorithms such as unrolled networks (50–53) which can learn more efficient ways to perform gradient descent along with learned regularizers.

## *Conclusion*

Wave-CAIPI was proposed to enable combined and high in-plane and SMS acceleration factors to improve the geometric fidelity and acquisition speed of EPI. Its application was demonstrated in GE, SE, and diffusion contrasts with single- and multi-shot acquisition, and g-Slider RF encoding was incorporated for high isotropic resolution imaging. Dual wave-PSFs and structured low-rank regularization were introduced to improve the robustness of its reconstruction to trajectory errors and shot-to-shot variations. We proposed joint image reconstruction method for wave-EPI with g-Slider to minimize the intra-voxel blurring by wave-encoding. In vivo experiments showed that wave-EPI acquisition allows high acceleration rates with decreased g-factor penalty and image artifacts relative to blipped-CAIPI EPI.

## *Acknowledgment*

This work was supported in part by NIH research grants: R01EB028797, R01MH116173, U01EB025162, P41EB030006, U01EB026996, R03EB031175 and the shared instrumentation grants: S10RR023401, S10RR019307, S10RR019254, and S10RR023043.

## *Captions*

**Figure 1**. **a**. The proposed wave-EPI based on SE-EPI. Half-cycle of cosine wave-encoding and one-cycle of sine wave-encoding were applied in the phase-encoding and the slice-selection directions, respectively. **b**. wave-EPI trajectory in k-space **c**. the concept of the wave-encoding deconvolution using the dual wave-PSFs.

**Figure 2**. The reconstructed images, g-factor analysis, and tSNR maps of single-shot GE-EPI at $R_{in}xR_{sms}$=3x3.

**Figure 3**. 32 diffusion-direction DWI with 1000s/mm$^2$ of b-value using the two-shot SE-EPI at $R_{in}xR_{sms}$=5x2 per each EPI-shot.

**Figure 4**. 32 diffusion-direction colored FA maps with 1000s/mm$^2$ of b-value using the two-shot SE-EPI at $R_{in}xR_{sms}$=5x2 per each EPI-shot.

**Figure 5**. 63 diffusion-direction DWI with 1000s/mm$^2$ of b-value using the two-shot SE-EPI at $R_{in}xR_{sms}$=6x2 per each EPI-shot. Five different encoded RFs were used for high-resolution g-Slider. We reconstructed wave-EPI images using standard wave-EPI reconstruction before g-Slider decoding and joint reconstruction of wave-EPI and g-Slider

**Figure 6**. 63 diffusion-direction colored FA maps with 1000s/mm$^2$ of b-value using the two-shot SE-EPI at $R_{in}xR_{sms}$=6x2 per each EPI-shot. Five different encoded RFs were used for high-resolution g-Slider.

**Figure 7**. Signal spreading simulation at 1.25mm isotropy voxel size with $R_{in}xR_{sms}$=4x3. One- and half-cycle cosine wave-gradients had 15mT/m and 30mT/m amplitudes in the phase-encoding direction, respectively, and 15mT/m of one-cycle sine wave-gradient was applied in the slice-selection direction in both cases. The stimulation levels for one TR were shown in the last column, which were computed by SIEMENS IDEA software for pulse sequence programming.

**Figure 8**. Intra-voxel dephasing simulation at 1mm isotropy voxel size. 22mT/m x 19mT/m of wave-encoding were applied. **a**. the phase variation within a voxel **b**. wave gradients and corresponding spatial frequencies during a readout **c**. PSF simulations using the standard wave-EPI reconstruction and joint reconstruction of wave-EPI and g-Slider.

Table 1. The imaging parameters for in-vivo experiments.

Supporting Figure S1. The number of $2^{nd}$-order crossing fibers according to the number of the keeping singular values during the SVD truncation for the low-rank regularization.

Supporting Figure S2. The reconstructed images, g-factor analysis, and tSNR maps of single-shot GE-EPI at $R_{in}$x$R_{sms}$=4x3.

Supporting Figure S3. Single diffusion direction image with 1000s/mm$^2$ of b-value using the two-shot SE-EPI at $R_{in}$x$R_{sms}$=5x2 per each EPI-shot.

## Figures

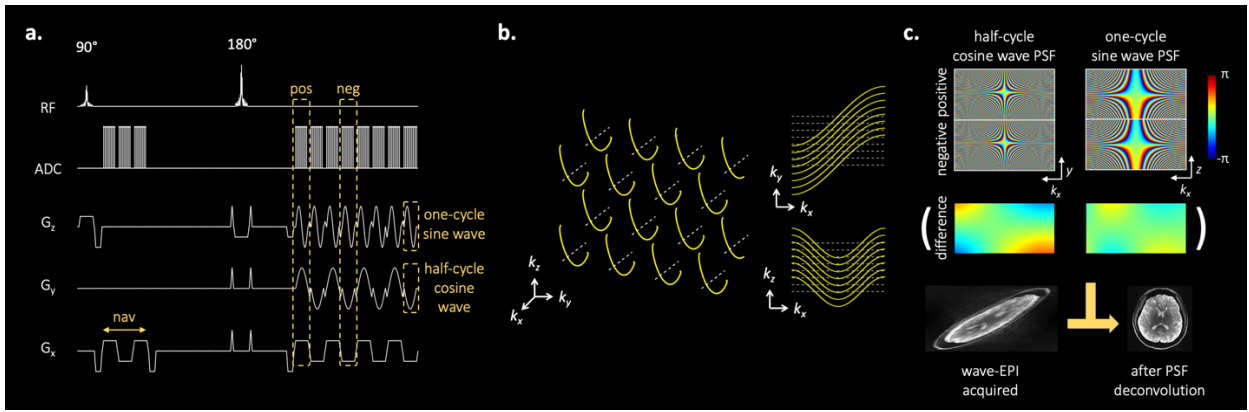

**Figure 1**. **a**. The proposed wave-EPI based on SE-EPI. Half-cycle of cosine wave-encoding and one-cycle of sine wave-encoding were applied in the phase-encoding and the slice-selection directions, respectively. **b**. wave-EPI trajectory in k-space **c**. the concept of the wave-encoding deconvolution using the dual wave-PSFs.

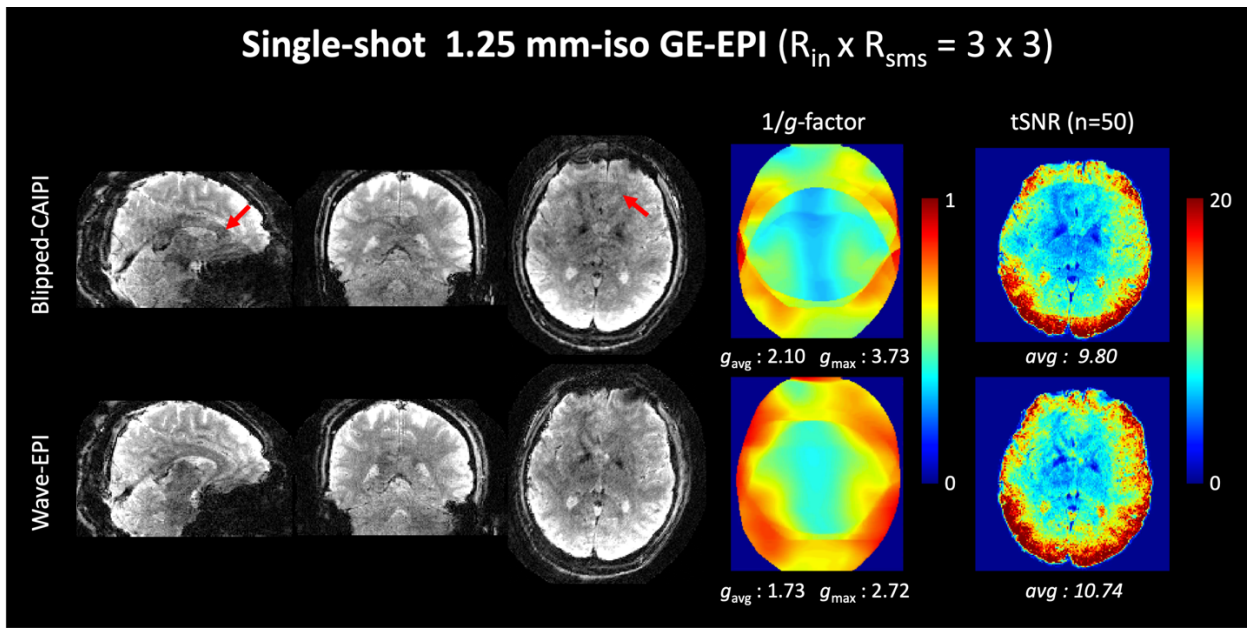

**Figure 2**. The reconstructed images, g-factor analysis, and tSNR maps of single-shot GE-EPI at $R_{in}×R_{sms}$=3x3.

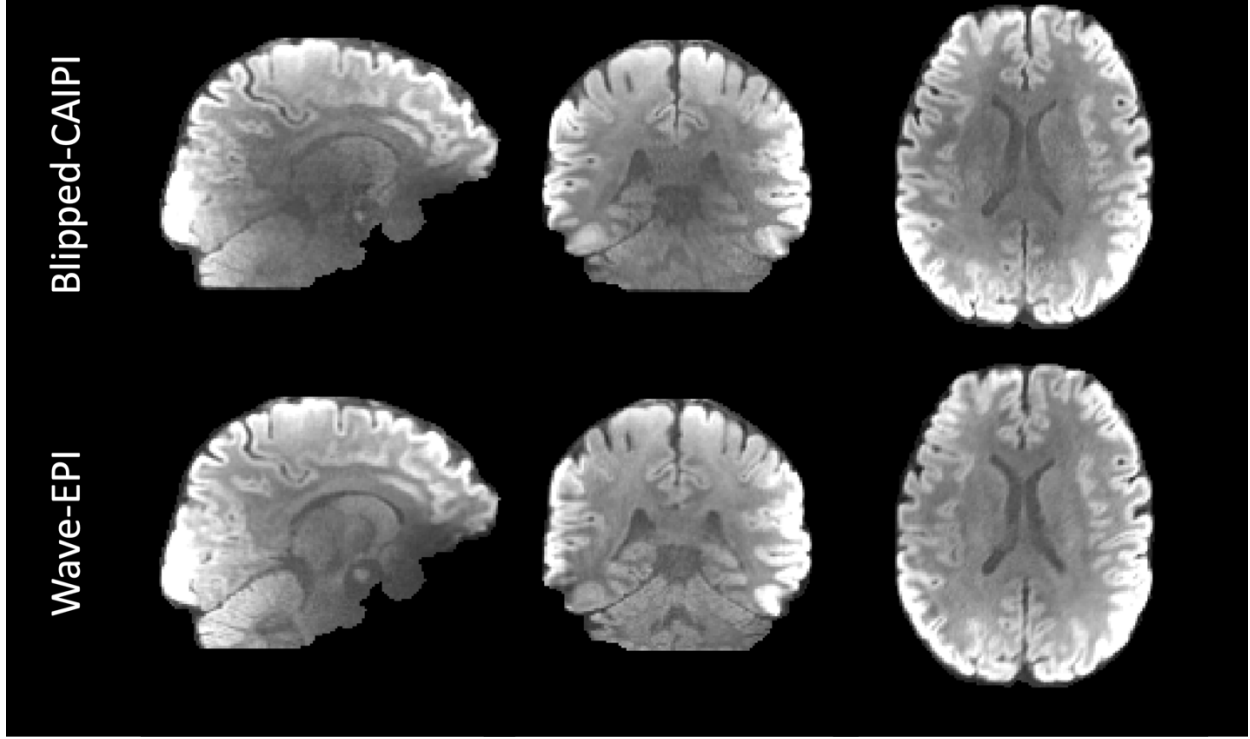

Figure 3. 32 diffusion-direction DWI with 1000s/mm$^2$ of b-value using the two-shot SE-EPI at $R_{in} \times R_{sms}$=5x2 per each EPI-shot.

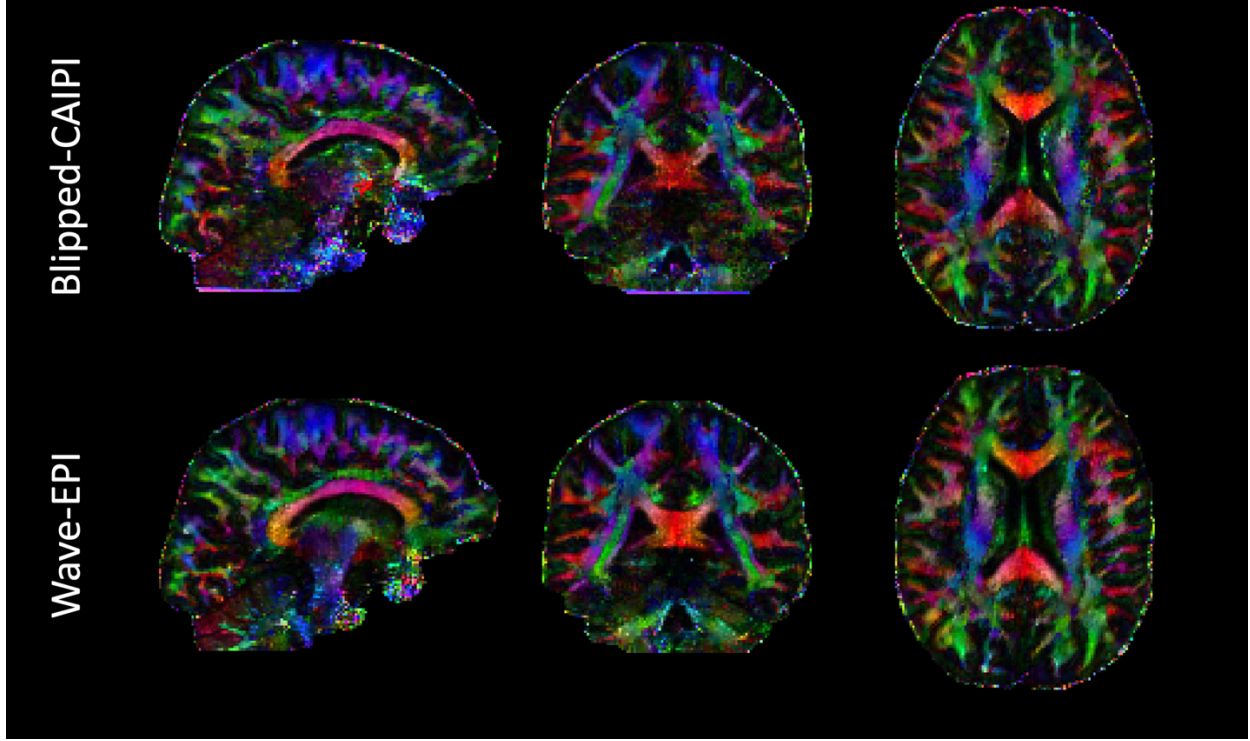

Figure 4. 32 diffusion-direction colored FA maps with 1000s/mm$^2$ of b-value using the two-shot SE-EPI at $R_{in}$x$R_{sms}$=5x2 per each EPI-shot.

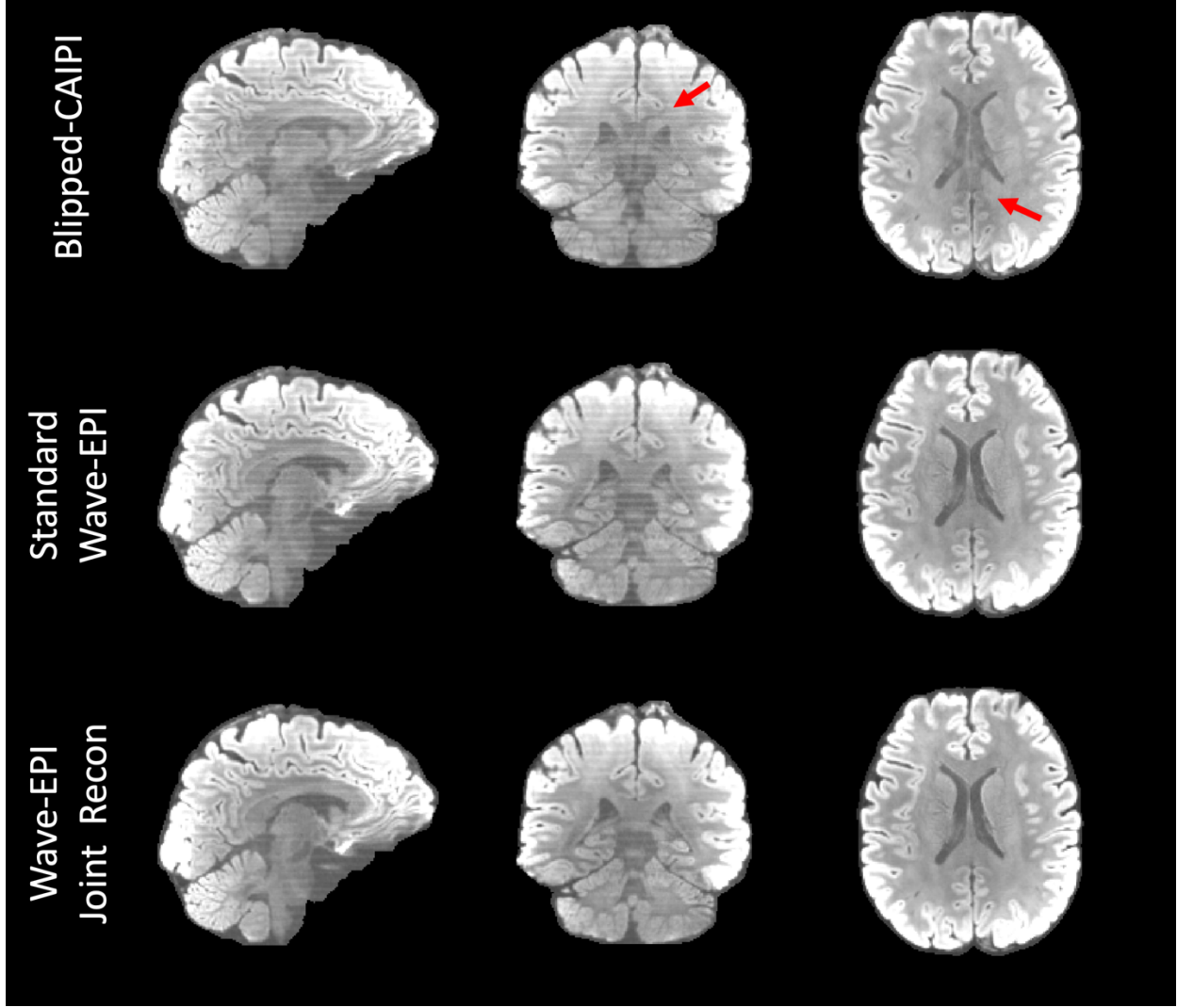

Figure 5. 63 diffusion-direction DWI with 1000s/mm$^2$ of b-value using the two-shot SE-EPI at $R_{in}$x$R_{sms}$=6x2 per each EPI-shot. Five different encoded RFs were used for high-resolution g-Slider. We reconstructed wave-EPI images using standard wave-EPI reconstruction before g-Slider decoding and joint reconstruction of wave-EPI and g-Slider

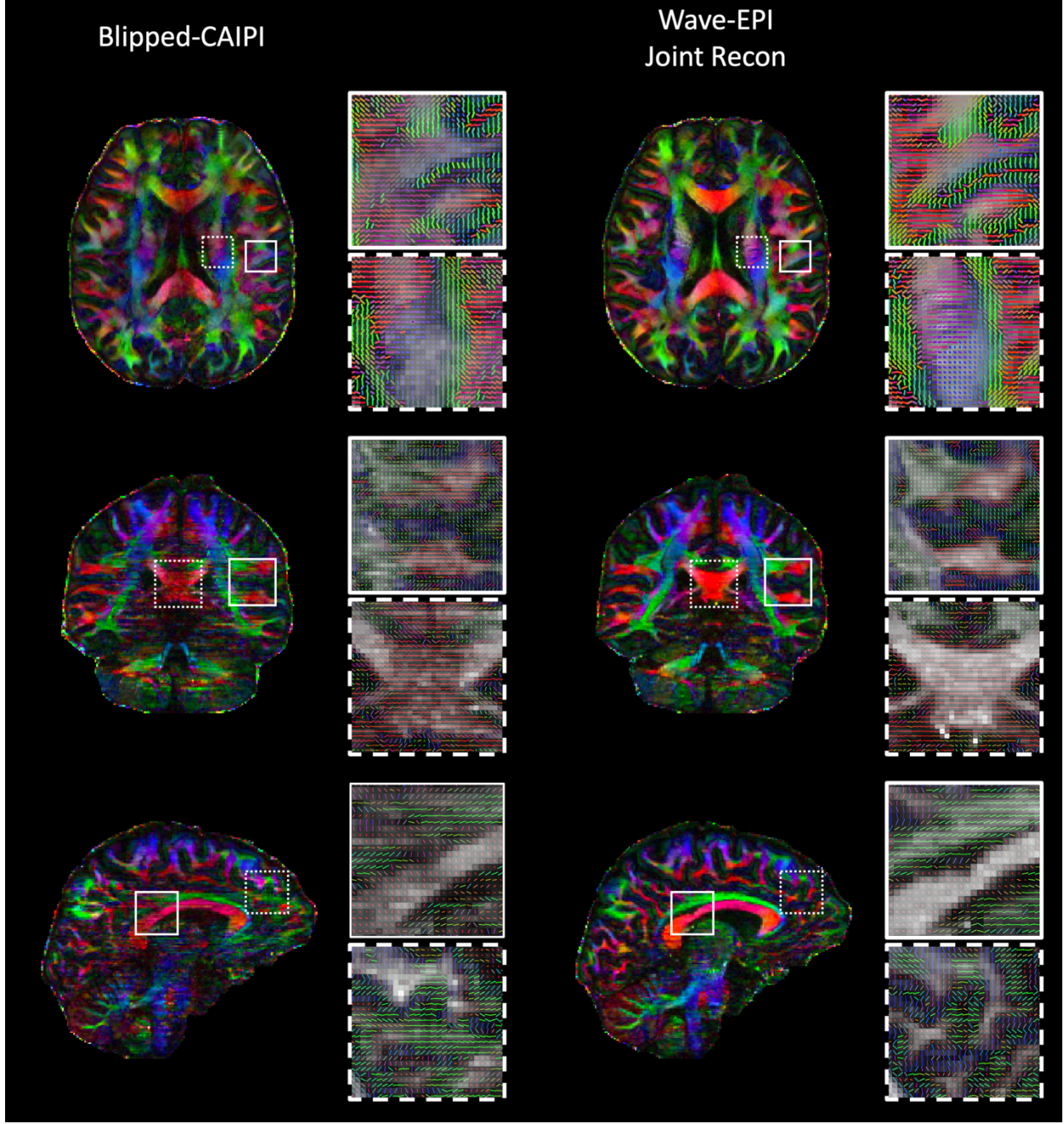

Figure 6. 63 diffusion-direction colored FA maps with 1000s/mm$^2$ of b-value using the two-shot SE-EPI at $R_{in} \times R_{sms}$=6x2 per each EPI-shot. Five different encoded RFs were used for high-resolution g-Slider.

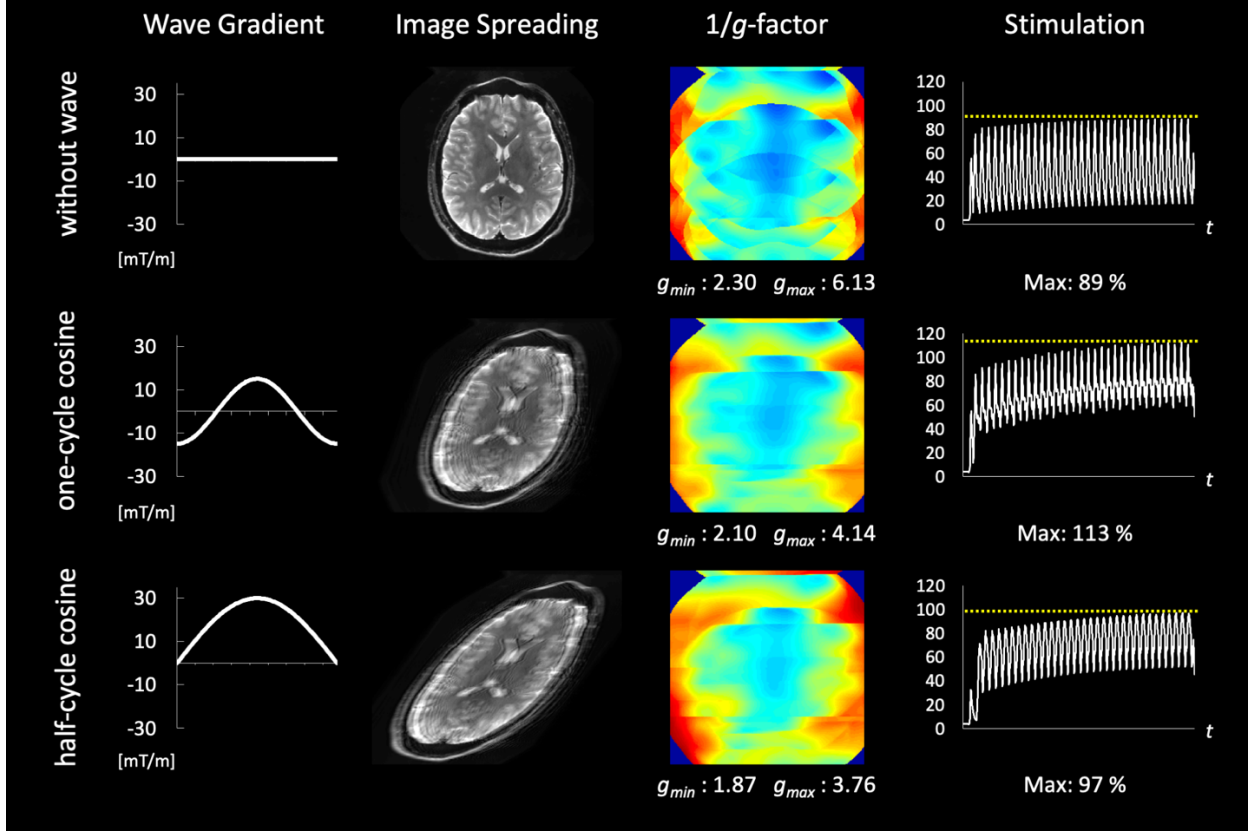

**Figure 7**. Signal spreading simulation at 1.25mm isotropy voxel size with $R_{in} \times R_{sms}$=4x3. One- and half-cycle cosine wave-gradients had 15mT/m and 30mT/m amplitudes in the phase-encoding direction, respectively, and 15mT/m of one-cycle sine wave-gradient was applied in the slice-selection direction in both cases. The stimulation levels for one TR were shown in the last column, which were computed by SIEMENS IDEA software for pulse sequence programming.

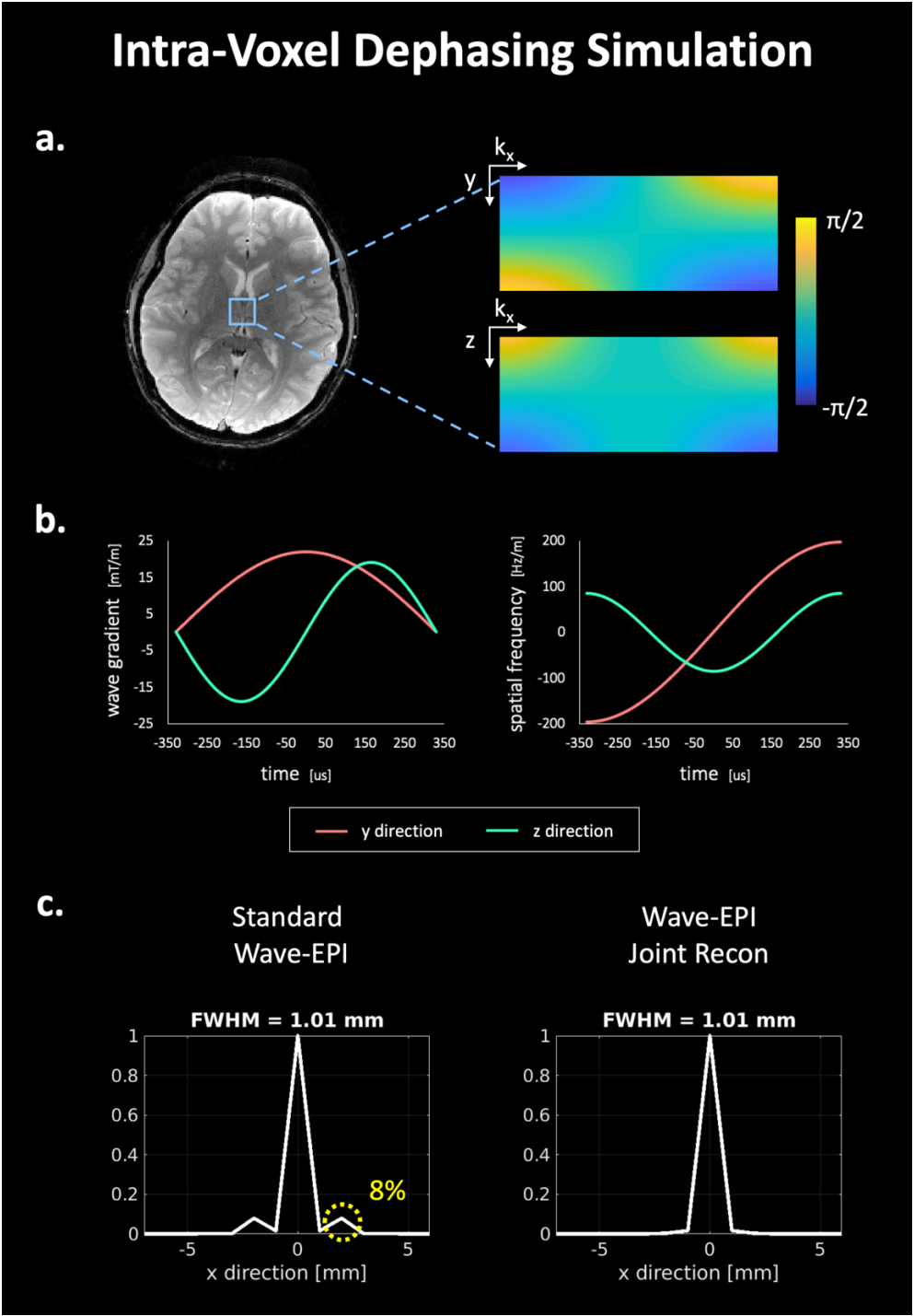

Figure 8. Intra-voxel dephasing simulation at 1mm isotropy voxel size. 22mT/m x 19mT/m of wave-encoding were applied. a. the phase variation within a voxel b. wave gradients and corresponding spatial frequencies during a readout c. PSF simulations using the standard wave-EPI reconstruction and joint reconstruction of wave-EPI and g-Slider.

**Table 1.** The imaging parameters for in-vivo experiments.

| | GE-EPI | | GE-EPI | | dMRI SE-EPI | | dMRI g-Slider | |
| --- | --- | --- | --- | --- | --- | --- | --- | --- |
| | Figure 2 | | Figure S2 | | Figure 3, 4, S3 | | Figure 5, 6, S1 | |
| | Blipped-CAIPI | Wave-EPI | Blipped-CAIPI | Wave-EPI | Blipped-CAIPI | Wave-EPI | Blipped-CAIPI | Wave-EPI |
| Imaging Plane | Axial | | Axial | | Axial | | Axial | |
| FOV [mm$^3$] | 220 x 220 x 132 | | 220 x 220 x 132 | | 220 x 220 x 130 | | 220 x 216 x 150 | |
| Resolution [mm$^3$] | 1.25 x 1.25 x 1.25 | | 1.25 x 1.25 x 1.25 | | 1.25 x 1.25 x 1.25 | | 1.00 x 1.00 x 1.00 | |
| TR [ms] | 2800 | | 2400 | | 6900 | | 3000 | |
| TE [ms] | 27 | 28 | 23 | 24 | 73 | 73 | 70 | 71 |
| Wave-gradient [mT/m]$^2$ | - | 29 x 15 | - | 29 x 15 | - | 30 x 15 | - | 22 x 19 |
| # of shot | 1 | | 1 | | 2 | | 2 | |
| Reduction factor per each shot ($R_{shot} = R_{in} \times R_{sms}$) | 3 x 3 | | 4 x 3 | | 5 x 2 | | 6 x 2 | |
| Receiver bandwidth | 1672 | | 1672 | | 1672 | | 1516 | |
| Partial Fourier factor | 6 / 8 | | 6 / 8 | | 6 / 8 | | 8 / 8 | |
| b-value [s/mm$^2$] | - | | - | | 1000 | | 1000 | |
| # of diffusion directions | - | | - | | 32 | | 63 | |
| # of b0 acquisition | - | | - | | 4 | | 8 | |
| # of RFs for g-Slider encoding | - | | - | | - | | 5 | |

Table 1. The imaging parameters for in-vivo experiments.

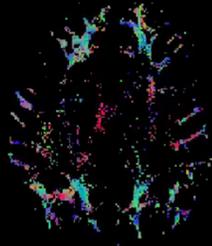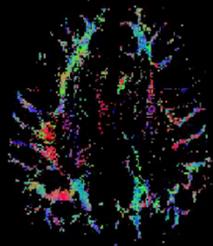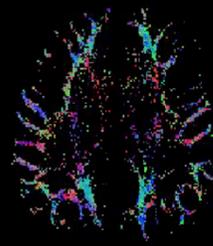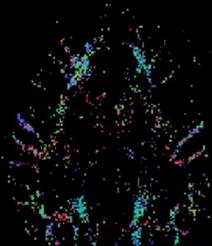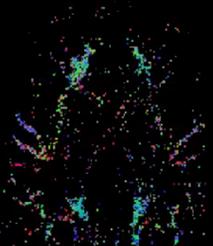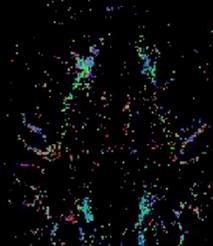

**Supporting Figure S1.** The number of 2nd-order crossing fibers according to the number of the keeping singular values during the SVD truncation for the low-rank regularization.

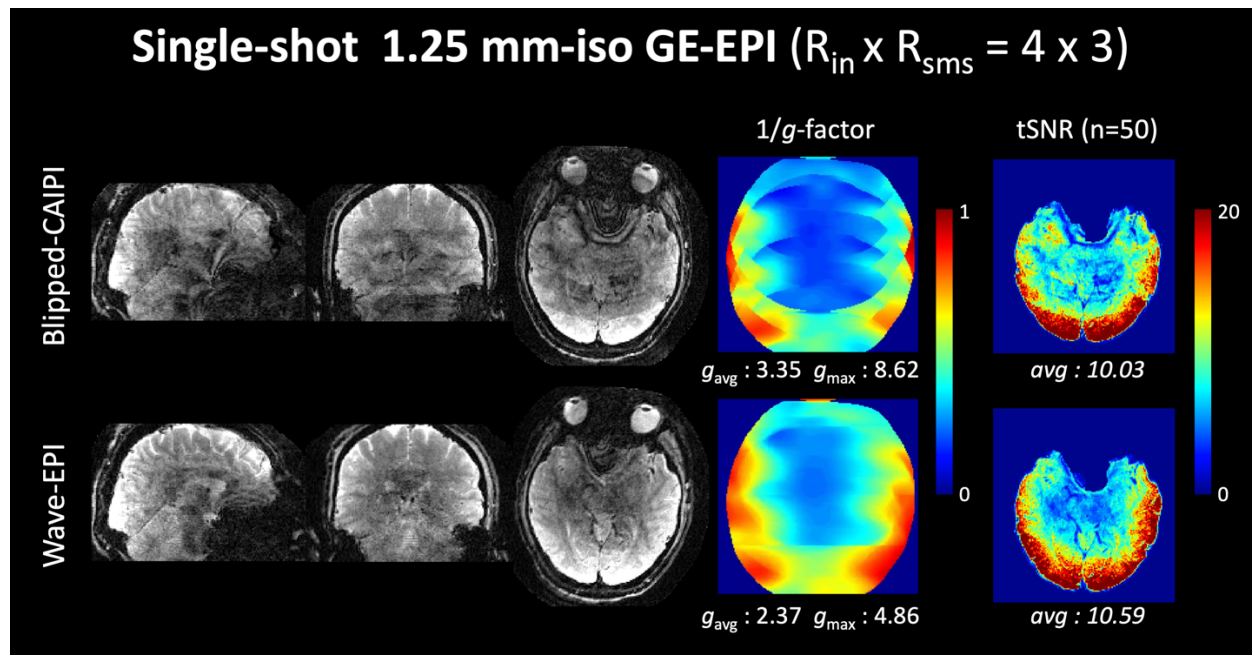

Supporting Figure S2. The reconstructed images, g-factor analysis, and tSNR maps of single-shot GE-EPI at $R_{in} \times R_{sms}$=4x3.

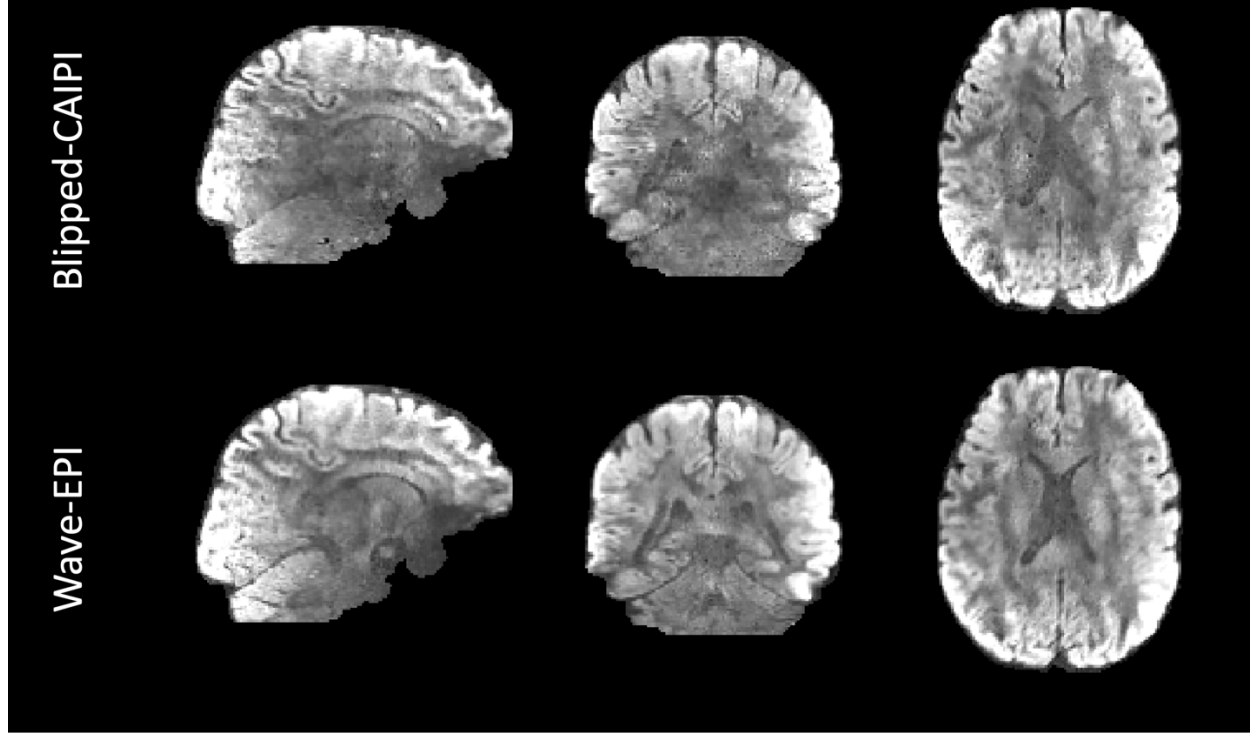

**Supporting Figure S3**. Single diffusion direction image with 1000s/mm$^2$ of b-value using the two-shot SE-EPI at $R_{in}xR_{sms}$=5x2 per each EPI-shot.